\documentclass[twocolumn,aps,prd,showpacs]{revtex4}
\usepackage{graphicx}
\usepackage{amsmath}
\usepackage{bm}

\begin{document}

\title{\mbox{}\\[10pt]
Factorization in Break-up and Recombination Processes\\
for Atoms with a Large Scattering Length}

\author{Eric Braaten and Dongqing Zhang}
\affiliation{
Physics Department, The Ohio State University, 
Columbus, Ohio 43210, USA}

\date{\today}
\begin{abstract}
Break-up and recombination processes for loosely-bound molecules
composed of atoms with a large scattering length $a$ 
necessarily involve interactions 
that are nonperturbative in the exact 2-body interaction.
If these processes involve atoms with relative momenta 
much larger than $\hbar/|a|$, the leading contributions to their
rates can be separated into short-distance factors that are 
insensitive to $a$ and long-distance factors that are insensitive
to the range of the interaction. 
These factorization contributions can be obtained 
from the leading term in a perturbation expansion 
in the exact atom-atom scattering amplitude. 
The short-distance factors 
are atom-atom cross sections at a lower collision energy.
In the special case of inclusive 
break-up cross sections for atom-molecule scattering,
the long-distance factors 
simply count the number of atoms in the molecule.
\end{abstract}



\maketitle


Nonrelativistic particles with short-range interactions that 
have been tuned, either by the experimenter or fortuitously 
by nature, so that their S-wave scattering length $a$ is much 
larger than the range, have universal low-energy properties 
that depend on $a$ but are otherwise insensitive to their
interactions at short distances.
(See Ref.~\cite{Braaten:2004rn} and references therein.) 
If the particles form loosely-bound 2-body
or higher $N$-body clusters whose sizes are comparable to $|a|$, 
the clusters also have  universal properties.
A classic example in atomic physics is $^{4}$He atoms whose
scattering length $a \approx 100$ \AA\ is much 
larger than their effective range $r_s \approx 7$ \AA. 
The $^{4}$He dimer and the excited state of the $^{4}$He trimer 
both have sizes comparable to $a$.  
A classic example in nuclear physics is nucleons.
The deuteron is a bound state of the neutron and proton associated 
with a large scattering length in the spin-triplet channel. 

The large scattering length implies that interactions 
between atoms whose relative momenta are comparable 
to $\hbar/|a|$ are nonperturbative in the 2-body scattering amplitude.
A simple consequence of these strong interactions with $a>0$
is the existence of a loosely-bound dimer
whose binding energy is given by
\begin{equation}
E_D = \hbar^2/(ma^2).
\label{Edimer}
\end{equation}
In some cases (for example, identical bosons), 
the strong interactions produce the {\it Efimov effect}:  
as $a \to \pm \infty$, there are increasingly 
many loosely-bound trimers with an accumulation point 
at the 3-atom threshold \cite{Efimov70}. 
In the limit $a = \pm \infty$, 
the trimers have an asymptotically exponential spectrum: 
\begin{equation}
E_T^{(n)} \longrightarrow (e^{-2 \pi/s_0})^{n-n_*} \hbar^2 \kappa_*^2/m
\qquad {\rm \ as \ } n \to \infty,
\label{Etrimer}
\end{equation}
where $s_0$ is a numerical constant whose value for identical bosons
is $1.00624$, $n_*$ is an arbitrary integer, and $\kappa_*$ is a 
3-body parameter \cite{Braaten:2004rn}.
The strong interactions can also lead to intricate dependence 
on the scattering length $a$. 
An example in the case of identical bosons is the event rate 
constant in the low-energy limit 
for 3-body recombination into the loosely-bound dimer:
\begin{equation}
K_3 = 
{768 \pi^2 (4 \pi - 3\sqrt{3}) \hbar a^4 /m\over 
	\sinh^2(\pi s_0) + \cosh^2(\pi s_0) \cot^2[ s_0 \ln (a \kappa_*)+ 1.16]} .
\label{alpha-0}
\end{equation}
The log-periodic dependence on $a$ in Eq.~(\ref{alpha-0}) 
was discovered in Refs.~\cite{alpha}.  The completely analytic expression 
was derived more recently by Petrov \cite{petrov}.
The coefficient of $\hbar a^4/m$ on the right side of Eq.~(\ref{alpha-0})
can range from 0 to 402.7 depending on the value of the 3-body 
parameter $\kappa_*$. 
The dimer break-up cross section in atom-dimer scattering
at a collision energy $E$ just above the threshold $E_D$
has the same log-periodic factor:
\begin{equation}
\sigma_{AD}^{\rm (break-up)}(E) = {\sqrt{3} \over 96 \pi} 
\left({m K_3 \over \hbar a^4} \right) 
a^2 {(E - E_D)^2 \over E^{1/2} E_D^{3/2}}.
\label{sigmaAD-0}
\end{equation}
To reproduce such intricate dependence on $a$ requires
accurate numerical methods that are capable of solving
the problem ``exactly'', that is, to any desired precision.
The Schr\"{o}dinger equation for 2 or 3 atoms interacting 
through a short-range potential can be solved exactly.
Systems of 4 atoms are at the frontiers of few-body physics:
the binding energies of tetramers can be 
calculated accurately, but there are some scattering 
observables for which effective calculational methods 
have yet to be developed.
In any process that involves loosely-bound molecules, 
strong interactions involving the momentum scale $\hbar/|a|$ are 
unavoidable. This suggests that accurate calculations of 
the rates for such processes may be possible using present methods
only if the total number of atoms involved is at most 3 or 4.

In this Letter, we point out that this obstacle can be avoided 
for processes involving atoms whose momenta relative to the 
loosely-bound molecules are set by a scale $Q \gg \hbar/|a|$. 
Examples of such processes are the break-up of the molecule 
in a collision with an energetic atom and the formation 
of the molecule in a collision involving energetic atoms.
The separation of scales $Q \gg \hbar/|a|$ can be exploited by using
{\it factorization} to separate the rate into
{\it short-distance} factors that are insensitive to $a$ and 
{\it long-distance} factors that do not involve the scale $Q$.
The strong interactions associated with large $a$
enter only in the long-distance factors.
The rate can be calculated accurately using present methods provided
the long-distance factors involve at most 3 or 4 atoms. 
In some cases, even this limitation is unnecessary
because the long-distance factors can be determined analytically.
Thus factorization significantly expands the list of processes 
whose rates can be calculated accurately using present methods.

Factorization has proved to be a powerful tool
in quantum chromodynamics (QCD). 
The Nobel Prize in physics for 2004 was awarded to 
Gross, Politzer, and Wilczek for the discovery of the
{\it asymptotic freedom} of QCD \cite{QCD}.
Asymptotic freedom 
refers to the decrease of the strength of the interaction between 
quarks as their separation decreases, or equivalently,
as their relative momentum increases. For selected observables,
asymptotic freedom can be exploited by using factorization 
to separate the observables into short-distance factors that 
involve only weak QCD interactions between quarks and gluons 
and long-distance factors that involve strong QCD interactions 
\cite{Collins:1987pm}. 
The short-distance factors involve a large momentum transfer $Q$ 
and can be calculated using perturbation theory in the 
running coupling constant $\alpha_s(Q)$ of QCD. 
The long-distance factors involve only momentum transfers
small compared to $Q$. In most cases, systematic methods for 
calculating the long-distance factors have not been developed. 
The bulk of the quantitative evidence that QCD describes the 
strong force is the experimental verification that the short-distance factors
are correctly predicted by perturbative QCD. 

Atoms with a large scattering length have a property analogous 
to asymptotic freedom: the strength of their interaction,
as measured by the magnitude of the elastic scattering amplitude,
decreases as their relative momentum $\hbar k$ increases.
This behavior of the atom-atom interaction motivates our use of the
factorization strategy that has been so successful in QCD. 
The S-wave elastic scattering amplitude has the form
\begin{equation}
f_k = 
\left[ (- 1/a + \mbox{${1 \over 2}$} r_s k^2 + \ldots) - i k \right]^{-1},
\label{fk}
\end{equation}
where $r_s$ is the effective range.
For collision energies $E$ at which higher partial waves 
can be neglected, the elastic cross section for identical bosons is
\begin{equation}
\sigma _{\rm AA} (E) = 
{8 \pi \over (-1/a + {1 \over 2} r_s m E/\hbar^2 + \ldots)^2 + mE/\hbar^2}.
\label{sigmaAA}
\end{equation}
The cross section has its maximum value $8 \pi a^2$ at $E=0$.
In the scaling region 
$\hbar^2/m a^2 \ll E \ll \hbar^2/m r_s^2$, 
the cross section scales like $1/E$:
\begin{equation}
\sigma _{\rm AA} (E) \approx
{8 \pi \hbar^2 \over mE}.
\label{sigmaAA:scaling}
\end{equation}
%

\begin{figure}
\includegraphics[width=6cm]{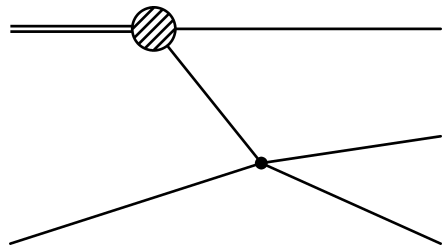}
\caption{Feynman diagram for the break-up of a dimer 
by the hard scattering of one of its constituent atoms 
and a colliding atom. The blob represents 
the wavefunction of the dimer and the dot represents 
the exact atom-atom scattering amplitude. 
\label{fig:ADtoAAA}}
\end{figure}

Consider the break-up of a loosely-bound dimer 
through a collision with an atom. 
We take all three atoms to be identical bosons. 
The two atoms inside the dimer have typical separation $a$.  
The atom-dimer collision energy is $E={3\over 4}Q^2/m$,
where $Q$ is the momentum of the atom or the dimer 
in the center-of-mass frame.
If $Q \gg \hbar/a$,
the leading contribution to the break-up of the dimer comes from 
one of its constituent atoms being knocked out by  
the energetic incoming atom, 
while the other acts as a spectator. 
This process can be represented by the Feynman diagram 
in Fig.~\ref{fig:ADtoAAA}, which is the leading term in a
perturbation expansion in the exact atom-atom scattering amplitude.
The expression for the T-matrix element is
\begin{equation}
{\cal T} = \sum_{(123)}
{(32 \pi \hbar^3/m)(\pi/a)^{1/2}(1 +\frac{3}{4} r_s/a + \ldots) 
\over (q_3^2 + m E_D)[(-1/a+{1 \over 2} r_s k_{12}^2 + \ldots) - ik_{12}]},
\label{T-LO}
\end{equation}
where $E_D = \hbar^2/(m a^2)(1 + r_s/a + \ldots)$
is the dimer binding energy including range corrections,
$q_3$ is the relative momentum of the two atoms in the dimer
(one of which is the outgoing atom labelled 3),
and $\hbar k_{12} = {3 \over 4} Q$ is the relative momentum of the 
incoming atom and the scattered atom from the dimer 
(which is also the relative momentum of the outgoing atoms
labelled 1 and 2). 
The sum is over cyclic permutations of the three outgoing atoms. 
The effects of subsequent interactions between the scattered atoms 
and the spectator atom in the dimer enter through higher-order diagrams
in the perturbation expansion 
in the exact atom-atom scattering amplitude.
These diagrams are all suppressed by a factor of $\hbar/(aQ)$ 
and can be neglected if the collision energy is
sufficiently high.  If we take the limit $Q \gg \hbar/a$,
the $T$-matrix element in Eq.~(\ref{T-LO})
can be expressed as the product of a
short-distance factor that involves only the large momentum scales 
$Q$ and $\hbar/r_s$ and a long-distance factor that involves only 
the small momentum scales $q$ and $\hbar/a$:
\begin{equation}
{\cal T} \approx  \sum_{(123)}
{4 \hbar^2 (\pi /a)^{1/2} \over q_3^2 + \hbar^2/a^2} \times
\: {8 \pi \hbar/m \over ({1 \over 2} r_s k_{12}^2 + \ldots) - i k_{12}}.
\label{T-fact}
\end{equation}
The long-distance factor is proportional to 
the momentum-space wavefunction of the dimer in the zero-range limit. 
The cross section is obtained by squaring the T-matrix element 
and integrating over the momenta of the three identical outgoing atoms.
The cross section is dominated by the 
squares of the terms in Eq.~(\ref{T-fact}), 
which we call diagonal terms.
The terms in the cross section that correspond 
to the interference between terms in Eq.~(\ref{T-fact}), 
which we call cross terms,
are suppressed by a power of $\hbar/(aQ)$.
We can take the limit $Q \gg \hbar/a$ in the momentum integrals. 
The result for the break-up cross section at collision 
energy $E \gg \hbar^2/ma^2$ can be expressed in the factored form
\begin{equation}
\sigma _{\rm AD} ^{(\rm break-up)} (E) \approx 
2 \, \sigma _{\rm AA} (\mbox{${3\over4}$} E) .
\label{sigmaAD}
\end{equation}
The coefficient $2$, which is the long-distance factor, is obtained
by using the normalization integral for the dimer wavefunction. 
The factorization form in Eq.~(\ref{sigmaAD}) applies 
at arbitrarily large $E$ provided $\sigma_{AA}$ 
is the total atom-atom cross section.
In the scaling region $E_D \ll E \ll \hbar^2/m r_s^2$,
this breakup cross section scales as $1/E$: 
\begin{equation}
\sigma _{\rm AD} ^{(\rm break-up)} (E) \approx
{64 \pi \hbar^2 \over 3 m E}.
\label{sigmaAD:scaling}
\end{equation}

The result can be generalized to the inclusive break-up
cross sections for Efimov trimers and other loosely-bound
molecules whose constituents all have separations of order 
$|a|$. The leading contribution comes 
from one of the atoms in the molecule being knocked out by a 
collision with the energetic incoming atom.  If the 
molecule contains $N$ loosely-bound atoms, 
the atom-molecule collision energy is $E ={N+1 \over 2N} Q^2/m$,
where $Q$ is the momentum in the center-of-mass frame.
The short-distance factor in the T-matrix element
is the same as in Eq.~(\ref{T-fact}), except that the
relative momentum is $\hbar k = {N+1 \over 2N} Q$.
For any specific final state, the long-distance factor may be 
a very complicated function of $a$ and the 3-body parameter 
$\kappa_*$ that appears in Eq.~(\ref{alpha-0}). It involves 
the wavefunction of the molecule and the wavefunction
of the $(N-1)$-atom system that remains after the hard scattering.
Some of those $N-1$ atoms may be bound into dimers or other clusters.  
However, if we sum over all these $(N-1)$-atom states, 
we can use completeness relations to show that
the long-distance factor is simply the number $N$ of atoms in the molecule.  
Our result for the inclusive 
break-up cross section as a function of the collision energy $E$ is
\begin{equation}
\sigma _{\rm AM}^{(\rm break-up)} (E) \approx
N \,  \sigma _{\rm AA}(\mbox{${N+1 \over 2N}$} E) .
\label{sigmaAM}
\end{equation}
This expression is valid when the collision energy satisfies
$E \gg \hbar^2/ma^2$ and $E \gg E_M$, where $E_M$ is the binding 
energy of the molecule with respect to the $N$-atom threshold.

The cross section for the break-up of $^4$He dimers in collisions 
with Xe atoms has been studied in Ref.~\cite{BGR96}
for collision energies $E$ ranging from 46 K to 348 K.
Since the Xe atom is so much heavier than a He atom,
each of the atoms in the He dimer carries approximately 
${1\over2}$  the collision energy $E$.
The analog of our factorization formula in Eq.~(\ref{sigmaAD})
predicts that if $E$ is much larger than the He dimer 
binding energy, which is about 1.6 mK, 
the He$_2$--Xe break-up cross section 
should be approximately 2 times the He--Xe cross section 
at collision energy ${1\over2}E$. 
This prediction agrees reasonably well with the VCC--RIOS approximation 
\cite{VCCRIOS} studied in Ref.~\cite{BGR96}.
By comparing the cross section $\sigma^{\rm VCC-RIOS}_{\rm tot}$
in Table 2 of Ref.~\cite{BGR96} with the appropriate interpolated 
values of $\sigma^{\rm el}_{\rm HeXe}$,
we can see that the difference is less than 
9\% in the energy range from 93 K to 348 K. 
In the Independent-Atoms model considered in Ref.~\cite{BGR96},
the total cross section is 2 times 
the He--Xe cross section at collision energy $E$ 
instead of ${1\over2}E$. 
This model underestimates the VCC-RIOS cross section by 
an amount that ranges from 11\% at 93 K to 30\% at 348 K.

Factorization can also be applied to the rate for 3-body 
recombination into the loosely-bound dimer when the three incoming 
atoms all have relative momenta much greater than $\hbar/a$.
The leading contribution comes from a hard scattering 
of two of the atoms that scatters one of them into a state 
with small momentum relative to the third atom, 
followed by the coalescence of those two atoms into a dimer.
The coalescence probability 
is substantial only if the relative momentum of the scattered atom
and the third atom is of order $\hbar/a$. 
At leading order in the perturbation expansion in the
exact atom-atom scattering amplitude, the T-matrix element
for this process 
can be expressed as the sum of three Feynman diagrams 
obtained by time-reversing the diagram in Fig.~\ref{fig:ADtoAAA}
and summing over cyclic permutations of the three incoming atoms.
We work in the center-of-mass frame,
taking the momenta of the three incoming atoms to be
$\vec{p}_1$, $\vec{p}_2$, and $\vec{p}_3$ and the momenta of the 
outgoing dimer and atom to be $+\vec{Q}$ and $-\vec{Q}$.
The collision energy is $E=(p_1^2 + p_2^2 + p_3^2)/(2m)$.
For large collision energy $E$,
the leading contribution to the T-matrix element
is given by Eq.~(\ref{T-fact}).
Each of the three terms in the sum
is the product of a short-distance factor and
a long-distance factor.  The hard 
scattering of atoms $1$ and $2$ followed by the coalescence of 
one of them with atom 3 gives 
the term shown explicitly in Eq.~(\ref{T-fact}), 
where $\hbar \vec{k}_{12} = {1\over2}(\vec{p}_1 - \vec{p}_2)$
is the large relative momentum of atoms 1 and 2 
and $\vec{q}_3 = \vec{p}_3 -{1\over2}\vec{Q}$
is the small relative momentum between 
atom 3 and the atom it coalesces with. 
The other two terms are obtained by 
cyclic permutations of 1, 2, and 3.
The 3-body recombination rate is obtained by 
squaring the T-matrix element and integrating over 
the final momenta of the outgoing atom and dimer. 
In the limit $E \gg \hbar^2/(m a^2)$,
the rate $R$ for forming dimers per volume and per time is
\begin{equation}
R \approx
{256 \pi^2 \hbar^3 \over m^2}  \,
\sigma_{AA}(\mbox{$3\over4$}E) \,
\sum_{(123)} \delta(E - 3 p_3^2/m).
\label{R-3br}
\end{equation}
The three terms in the sum are diagonal terms that come from 
the squares of the terms in Eq.~(\ref{T-fact}).
The cross terms corresponding to the interference between 
terms in Eq.~(\ref{T-fact})
have been neglected because they are suppressed 
by a factor of $(E_{D} / E)^{1/2}$.

The event rate $R$ in Eq.~(\ref{R-3br}) can be reduced to a function 
$K_3$ of the collision energy $E$ by 
averaging over the momentum hyperangle $\alpha_3$ defined by 
$p_k = ({4\over3} m E)^{1/2} \cos \alpha_k$
and over the angle $\beta_3$ between the vectors
$\vec p_3$ and $\vec p_1 - \vec p_2$.
The hyperangular average of $R$ can be expressed as
\begin{equation}
\langle R \rangle = 
{2 \over \pi} \int_0^{\pi/2} d \alpha_3 \, \sin^2(2 \alpha_3) 
\int_0^\pi d \beta_3 \, \sin \beta_3 \, R.
\label{hyper}
\end{equation}
The hyperangular average of the rate in Eq.~(\ref{R-3br}) is 
\begin{eqnarray}
K_3 \equiv  \langle R \rangle \approx
{384 \sqrt{3} \pi \hbar^3 \over m^2 E} \sigma_{AA}(\mbox{$3\over4$}E).
\label{R-fact}
\end{eqnarray}
This result is valid when $E \gg E_D$.
In the scaling region $E_D \ll E \ll \hbar^2/m r_s^2$,
the rate in Eq.~(\ref{R-fact}) scales as $1/E^2$:
\begin{eqnarray}
K_3 \approx
{12288 \sqrt{3} \pi^2 \hbar^5 \over 3 m^3 E^2} .
\label{R-scaling}
\end{eqnarray}
%

\begin{figure}
\includegraphics[width=7cm,angle=270]{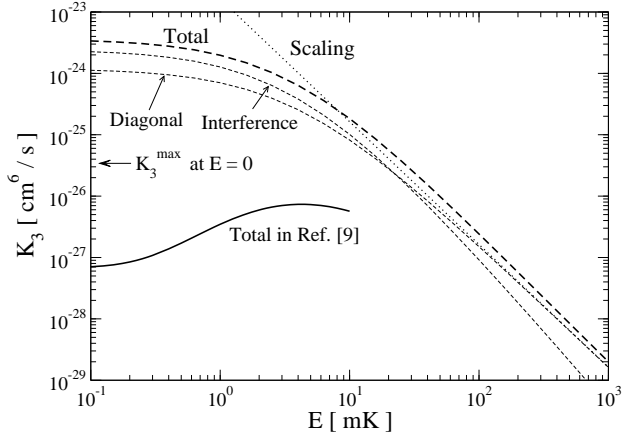}
\caption{Three-body recombination event rate $K_3$
(in units of cm$^6$/s) as a function of the collision energy $E$
(in units of mK).
The solid lines are the total result in Fig.~3 of Ref.~\cite{SEGB02}.
The dashed lines are the hyperspherical averages of the 
$R$ in Eq.~(\ref{R-LO}),
the diagonal terms, and the cross terms. 
The dotted line is the scaling approximation in Eq.~(\ref{R-scaling}).
The horizontal arrow indicates the maximum possible value of $K_3$ 
at $E=0$ allowed by Eq.~(\ref{alpha-0}).
\label{fig:aveR}}
\end{figure}

Suno et al.\ have calculated the 3-body recombination rate of 
$^{4}$He atoms into the dimer at collision energies up to 10 mK 
using accurate solutions of the 3-body Schr\"{o}dinger 
equation for the HFD-B3-FCI1 potential model \cite{SEGB02}.
In this model, the scattering length is $a=91.0$ \AA,
the effective range is $r_s \approx 7$ \AA,
and the dimer binding energy is $E_D = 1.6$ mK.
The scattering length deduced from the universal formula 
for the dimer binding energy in Eq.~(\ref{Edimer}) is $a_D =87.0$ \AA.
In Fig.~\ref{fig:aveR}, we compare 
the result of Ref.~\cite{SEGB02} for $K_3$
with the scaling approximation in Eq.~(\ref{R-scaling}). 
At $E=10$ mK, the highest collision energy 
considered in Ref.~\cite{SEGB02}, our scaling approximation 
in Eq.~(\ref{R-scaling})
is larger than the total result in Ref.~\cite{SEGB02} by a factor of 
28.8.  Our factorization approximation in Eq.~(\ref{R-fact})
is larger than the total result in Ref.~\cite{SEGB02} by a factor of 
23.7. 

We now consider why the scaling approximation in Eq.~(\ref{R-scaling}) 
overestimates the 3-body recombination rate at 10 mK 
by more than an order of magnitude.
We will show that the scaling approximation can be accurate 
only for collision energies that are at least an order of magnitude 
greater than 20 mK.
We calculate the sum of the contributions to $R$ at leading order 
in the perturbation expansion in the exact atom-atom scattering 
amplitude without making the factorization approximations.
The T-matrix element is given by Eq.~(\ref{T-LO}).
We set the effective range $r_s$ to zero to make the scaling behavior 
of the individual contributions at large energy $E$ more evident.
The resulting expression for the 3-body recombination rate is
\begin{widetext}
\begin{eqnarray}
R &=& {32 \sqrt{3} \pi \hbar^3 \over m^2 E^2} [E_D (E+E_D)]^{1/2} 
\sum_{(123)}
\bigg( \sigma_{AA}(E \sin^2 \alpha_3) 
\Big[ (\cos^2 \alpha_3 - \mbox{${1\over 4}$} 
	+ \mbox{${1\over 2}$} E_D/E)^2 
+ \mbox{${3\over 4}$} (E_D/E) (1+E_D/E) \Big]^{-1}
\nonumber
\\
&&  
+ 2 
{\sigma_{AA}(E \sin^2 \alpha_1) \sigma_{AA}(E \sin^2 \alpha_2)
\over \sigma_{AA}(E \sin \alpha_1 \sin \alpha_2)}
\nonumber
\\
&&  \hspace{0.5cm}
\times\int_0^1 dt 
\Big[ (\cos^2 \alpha_{12}(t) - \mbox{${1\over 4}$} 
	+ \mbox{${1\over 2}$} E_D/E)^2 
+ \mbox{${3\over 4}$} (E_D/E) (1+E_D/E)
+ 3 t(1-t) (1+E_D/E) \sin^2 \alpha_3 \Big]^{-1} \bigg).
\label{R-LO}
\end{eqnarray}
\end{widetext}
The angle $\alpha_{12}(t)$ is defined by
\begin{equation}
\cos^2 \alpha_{12}(t) = t \cos^2 \alpha_1 + (1-t) \cos^2 \alpha_2.
\end{equation}
The two terms inside the sum in Eq.~(\ref{R-LO})
are a diagonal term, which corresponds to the square of a
term in Eq.~(\ref{T-LO}),
and a cross term, which corresponds to the interference 
between two terms in Eq.~(\ref{T-LO}). 
In the diagonal term, the last factor peaks at 
$\cos \alpha_3 = {1\over4} - {1\over2}E_D/E$ with a width proportional to 
$[(E_D/E)(1+E_D/E)]^{1/2}$.  In the limit $E \gg E_D$, 
this factor can be approximated by a
delta function $\delta(\cos^2 \alpha_3 - \mbox{$1\over4$})$
multiplied by $(2\pi/\sqrt{3})(E/E_D)^{1/2}$.  The resulting expression 
for $R$ agrees with that in Eq.~(\ref{R-3br}).
The hyperangles $\alpha_1$ and $\alpha_2$ 
in Eq.~(\ref{R-LO}) can be expressed 
as functions of $\alpha_3$ and the angle $\beta_3$ between
the vectors $\vec p_3$ and $\vec p_1 - \vec p_2$:
\begin{eqnarray}
\cos^2 \alpha_{1,2} &=& \mbox{$1\over4$} \cos^2 \alpha_3 
+ \mbox{$3\over4$} \sin^2 \alpha_3 
\nonumber
\\
&& \mp  \mbox{$1\over2$}\sqrt{3} \cos \alpha_3 \sin \alpha_3 \cos \beta_3.
\end{eqnarray}
The hyperangular average of the rate $R$ in Eq.~(\ref{R-LO}) can then be 
calculated using Eq.~(\ref{hyper}).
In Fig.~\ref{fig:aveR}, we show $K_3 = \langle R \rangle$
as a function of the collision energy $E$, as well as
the contributions to $K_3$ from the diagonal terms and from 
the cross terms.
The cross terms are larger than the diagonal terms 
at low energies and smaller at higher energies,
with the crossover occurring near $E = 20$ mK.
At very high energy, the cross terms
scale like $E^{-5/2}$.  They eventually become negligible
compared to the diagonal terms,
which scale like $E^{-2}$.
The factorization approximation in Eq.~(\ref{R-fact})
is a high energy approximation to the 
contribution from the diagonal terms. 
From the crossover point and the scaling behavior,
we can infer that the factorization approximation 
can be a good approximation only if the collision energy exceeds the
20 mK by more than an order of magnitude.

\begin{figure}
\includegraphics[width=7cm,angle=270]{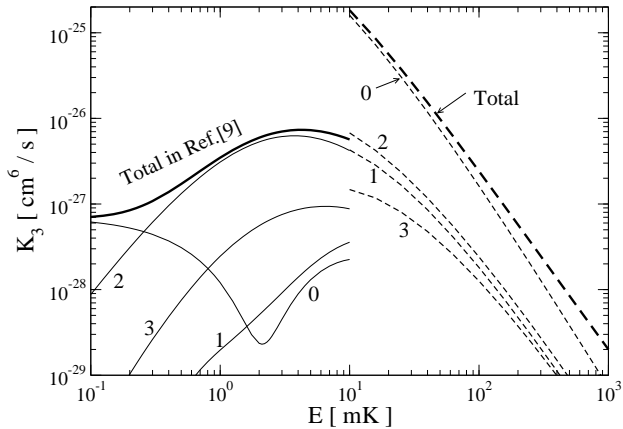}
\caption{Three-body recombination event rate $K_3$
(in units of cm$^6$/s) as a function of the collision energy $E$
(in units of mK).
The solid lines are the results in Fig.~3 of Ref.~\cite{SEGB02}.
The dashed lines are the hyperspherical averages of the $R$ 
in Eq.~(\ref{R-LO}) and the $J=0$, 1, 2 and 3 terms in Eq.~(\ref{R-amd}).  
\label{fig:Ramd}}
\end{figure}

We can get further insights into the factorization approximation 
by considering the angular-momentum decomposition of the 
3-body recombination rate.
The rate calculated by Suno et al.\ in Ref.~\cite{SEGB02} 
was obtained by adding
the contributions with total angular momentum quantum number
$J=0$, 1, 2 and 3.
These contributions are shown in Fig.~\ref{fig:Ramd}.
Note that at $E = 10$ mK, the highest energy for which they 
were calculated, the $J=0$ and 1 contributions are increasing
and they are smaller than the $J=2$ and 3 terms which are decreasing.
The angular-momentum decomposition of the rate in 
Eq.~(\ref{R-LO}) is
\begin{widetext}
\begin{eqnarray}
R
&=& {32\sqrt{3} \pi \hbar^3 \over m^2 E} \left( {E_D \over E+ E_D} \right)^{1/2}
\sum_{J=0}^\infty (2J+1)
\sum_{(123)} 
\Bigg( \sigma_{AA}(E \sin^2 \alpha_3) 
{Q_J(z_3)^2 \over \cos^2 \alpha_3} 
\nonumber
\\
&& \hspace{4.5cm}
+ 2 
{\sigma_{AA}(E \sin^2 \alpha_1) \sigma_{AA}(E \sin^2 \alpha_2)
\over \sigma_{AA}(E \sin \alpha_1 \sin \alpha_2)}
{Q_J(z_1) Q_J(z_2) \over \cos \alpha_1 \cos \alpha_2}
P_J(z_{3,12}) \Bigg),
\label{R-amd}
\end{eqnarray}
\end{widetext}
where the $Q_J(z)$ are Legendre functions of the second kind 
with branch cuts on the interval $-1 < z < +1$.  Their arguments are
\begin{equation}
z_i = {\cos^2 \alpha_i  + {1\over4} + E_D/E
	\over \sqrt{1+E_D/E} \cos \alpha_i}.
\end{equation}
The argument of the Legendre polynomial $P_J(z)$ in Eq.~(\ref{R-amd}) is
\begin{equation}
z_{3,12} = {\cos^2 \alpha_3 - \cos^2 \alpha_1 - \cos^2 \alpha_2
	\over 2 \cos \alpha_1 \cos \alpha_2}.
\end{equation}
The individual contributions from $J=0$, 1, 2, and 3 are shown in 
Fig.~\ref{fig:Ramd}.  In the high energy limit, 
each individual contribution scales like $E^{-5/2}$. 
Since the total $\langle R \rangle$ scales like $E^{-2}$,
the sum over $J$ must provide the additional factor of $E^{1/2}$.
At $E = 10$ mK, the $J=0$, 1, 2, and 3 terms
are larger than the results of Ref.~\cite{SEGB02} by factors of
711, 12.0, 1.6, and 1.7, respectively.  
The mismatch in the $J=2$ and 3 curves in Fig.~\ref{fig:Ramd} is small. 
The large mismatch in the 
$J=1$ curves is easy to understand.  The cross terms in 
Eq.~(\ref{R-amd}) are constructive for even $J$ and 
destructive for odd $J$, and they are 
much smaller than the diagonal terms for $J \ge 3$.  
Thus the $J=1$ channel is the only one with substantial 
destructive interference. 
The sum of the diagonal and cross terms is 
47\% of the diagonal terms at 100 mK, 22\% at 10 mK, and 4.5\% at 1 mK.
Because of the destructive interference at low energies, 
small corrections to the amplitude 
can give a large correction to the rate.
Thus only small corrections to the amplitudes for $J = 1$, 2, and 3
would be required for the terms in Eq.~(\ref{R-amd}) 
to match smoothly on to the results of Ref.~\cite{SEGB02}.
These corrections presumably come from higher orders in the 
perturbation expansion in the exact atom-atom scattering amplitude.

We now discuss the much larger mismatch between the $J=0$ curves in 
Fig.~\ref{fig:Ramd}.  
In this channel, there is constructive interference between 
the diagonal terms and the cross terms in the leading order result 
in Eq.~(\ref{R-amd}).
The corrections from higher orders in the exact atom-atom 
scattering ampitude could bring the curves into agreement if 
they interfere destructively with the leading order term. 
The destructive interference would have to decrease 
the $J=0$ term at $E=10$ mK by a factor of about 700. 
We argue that such strong destructive interference is plausible
by noting that there is strong destructive interference 
in the $J=0$ channel in the calculation of Ref.~\cite{SEGB02}.
The maximum possible value for $K_3$ at $E=0$ can
be obtained from Eq.~(\ref{alpha-0}) by setting $a_D = 87.0$ \AA.  
This value, $3.7\times 10^{-26}$ cm$^6$/s, is indicated by the
horizontal arrow in Fig.~\ref{fig:aveR}.
The value of $K_3$ at $E=0$ in Ref.~\cite{SEGB02} 
is smaller by a factor of 52,
which indicates strong destructive interference. 
The $J=0$ result in Ref.~\cite{SEGB02} exhibits even stronger 
destructive interference at an energy near 2 mK,
where it is smaller than at $E=0$ by a further factor of 31.
Thus the suppression from destructive interference at this energy 
is comparable to that required to make the $J=0$ curves
in Fig.~\ref{fig:Ramd} match smoothly at 10 mK. 
A nontrivial 3-body calculation would be required to
verify that higher orders in the exact atom-atom scattering amplitude
provide the necessary destructive interference. 
The result would depend not only on the scattering length $a$ or $a_D$, 
but also on the three-body parameter $\kappa_*$ that appears in 
Eqs.~(\ref{Etrimer}) and (\ref{alpha-0}).

The production of $^{4}$He dimers, trimers, and tetramers 
has been studied in experiments involving the free expansion 
of a jet of cold pressurized $^{4}$He atoms \cite{Bruch}.
The experiments are analyzed using coupled rate equations 
for various formation and break-up processes.  
The rate constants for processes involving dimers 
were estimated by using interpolations and extrapolations
of previous theoretical and experimental results.
Our results for dimer break-up in Eq.~(\ref{sigmaAD})
and for 3-body recombination in Eq.~(\ref{R-fact})
can be used to estimate the rates for those processes
at large collision energies.
The rate constants in Refs.~\cite{Bruch} for processes 
involving trimers were based on geometrical estimates only.  
Our result in Eq.~(\ref{sigmaAM}) for the inclusive break-up
cross section can be applied to both the excited state and the ground state
of the trimer.  Since the binding energy of the ground state trimer 
is about 50 times larger than that of the excited state,
the condition $E \gg E_T$ requires much larger collision energy
in the case of the ground state trimer.

Factorization can also be used to calculate the rate for 3-body 
recombination of two atoms and a loosely-bound dimer
into a loosely-bound trimer when the three incoming particles 
all have relative momenta much greater than $\hbar/a$.
The leading contribution comes from a hard scattering 
of the atoms that scatters one of them into a state with 
small momentum relative to the dimer, 
allowing it to subsequently coalesce with
the dimer to form a trimer.  The short-distance factor has a form 
similar to that in Eq.~(\ref{R-3br}).
The long-distance factor involves an overlap integral of
trimer and dimer wavefunctions.
The evaluation of this factor requires a nontrivial 
3-body calculation, but it is much simpler than the direct
calculation of the recombination rate by solving the 
4-atom Schr\"odinger equation. 
The factorization approximation could be used to calculate the
production rate of both the excited state and the ground state of
the $^4$He trimer in the experiment of Ref.~\cite{Bruch}.

The production of deuterons, $^3$He nuclei, and the corresponding 
antinuclei have been observed in ultrarelativistic heavy-ion 
collisions \cite{RHIC}.  These collisions are 
believed to produce a thermalized state with very high energy density
which, as it expands and cools, makes transitions to a 
quark-gluon plasma and then to a hadron gas.
Analogs of the factorization formulas in Eqs.~(\ref{sigmaAD}) and
(\ref{R-fact}) can be applied to the breakup and formation of 
deuterons and antideuterons in the hadron gas phase.

The strong interactions between atoms with large scattering lengths
implies that high numerical accuracy is required to calculate 
rates for processes involving loosely-bound molecules.
Using traditional methods, the list of processes for which
accurate calculations are possible is restricted to those involving 
at most 3 or 4 atoms.
By using factorization, the list can be expanded to include all
those for which the long-distance factors 
involve $N \le 3$ or 4 atoms.  In some cases, such as the
break-up cross sections in Eq.~(\ref{sigmaAM}),
the long-distance factors can be determined analytically 
and accurate calculations are possible even for $N > 4$. 

In the examples of $^4$He molecules and the deuteron,
the large scattering length $a$
arises from a fortuitous fine tuning by nature.  
Another exciting application of factorization
is to alkali atoms near a Feshbach resonance,
which allows $a$ to be tuned to arbitrarily 
large values by varying a magnetic field.
The factorization formulas give the leading
term in a systematic expansion in powers of $1/a$. 
As the scattering length is tuned to be increasingly large, 
the factorization approximation becomes increasingly accurate
and it applies at increasingly lower collision energies.

We acknowledge useful discussions with R.~Furnstahl.
We thank B.~Esry for providing us with the results of 
Ref.~\cite{SEGB02}. 
This research was supported in part by the Department of Energy
under grants DE-FG02-91ER4069 and DE-FG02-05ER15715. 


\end{document}